\documentstyle[prl,aps,multicol,epsf]{revtex}
\newcommand{\beq}{\begin{equation}}
\newcommand{\eeq}{\end{equation}}

\begin{document}
\begin{normalsize}

\title{Low-energy renormalization of the electron dispersion
of high-T$_c$ superconductors}

\author{{R. Zeyher}$^a$ and {A. Greco}$^{a,b}$} 

\address{$^a$ Max-Planck-Institut\  f\"ur\
Festk\"orperforschung,\\ Heisenbergstr.1, 70569 Stuttgart, Germany \\
$^b$Permanent address: Departamento de F\'{\i}sica, Facultad de
Ciencias Exactas e Ingenier\'{\i}a and \\
IFIR(UNR-CONICET), Av. Pellegrini 250, 
2000-Rosario, Argentina}

\date{\today}

\vspace{5cm}

\maketitle

\begin{abstract} 
High-resolution ARPES studies in cuprates
have detected low-energy changes in the dispersion and absorption of 
quasi-particles at low
temperatures, in particular, in the superconducting state. Based on
a new 1/N expansion of the t-J-Holstein model, which includes collective
antiferromagnetic fluctuations already in leading order,
we argue that the observed low-energy structures are
mainly caused by phonons and not by spin fluctuations, at least, in the
optimal and overdoped regime.

\end{abstract}

\newpage


ARPES studies in cuprates have recently detected changes in the electronic
spectral function near the Fermi energy 
at low temperatures and, in particular, in the superconducting state
\cite{Bogdanov,Valla,Kaminski,Lanzara}.
The dispersion of quasi-particles changes its slope by about a factor 2
over an energy region of about 50 meV 
and this change occurs in a rather isotropic
way around the Fermi surface. These results indicate that the particles 
near the Fermi surface 
interact with excitations characterized by an energy scale of about 50 meV
with a dimensionless coupling constant of about 1.
Possible candidates for these excitations are phonons or spin
fluctuations. In particular, the collective spin excitation associated with
the resonance peak of neutron scattering has been assumed to cause this
new low-energy scale\cite{Eschrig}.
 
Band structure calculations yield an overall dimensionless 
coupling constant $\lambda$ of electrons near the Fermi surface and phonons 
between 0.5 and 1.5\cite{Andersen,Krakauer}.
On the other hand, transport relaxation times 
indicate a very weak coupling $\lambda_{tr} < 0.3$ between electrons and 
phonons\cite{Tanner} in the optimally doped case. One explanation for this 
is that 
electronic correlations make the effective electron-phonon coupling strongly
momentum and frequency dependent suppressing 
large-momentum scattering\cite{Zeyher,Greco}. Below we will study  whether
a similar large reduction of the effective electron-phonon coupling 
by correlations occurs in the electronic self-energy and the
renormalization function $Z$ which determines the change in the quasi-particle
dispersion. Another relevant point is the anisotropy of this
renormalization. Assuming a bare, momentum-independent 
electron-phonon coupling the same correlation-induced momentum dependence
in the effective coupling which suppresses transport scattering rates 
will make $Z$ in general anisotropic. The calculation has to show whether
this induced anisotropy is small enough to account for the rather isotropic
behavior of the slope change observed in experiment.

Spin fluctuations within the usual 1/N expansion yield only small 
contributions to the electronic self-energy and to scattering rates at low
energies\cite{Greco}. Though the coupling between spin fluctuations and 
electrons
is set by the large hopping integral $t$,
the spin fluctuation spectrum at large N is rather structureless and extends
in energy over several $t's$. Most spin flips thus are high-energy 
excitations yielding only small renormalizations at low energies. 
Below we will present results from a different 1/N
expansion which favors antiferromagnetic fluctuations and accounts for 
the resonance
peak in the spin susceptibility already in leading order.
This approach is convenient to make a 
fair comparison of phonon- and spin fluctuations-induced
contributions to the electronic self-energy. 
 

In the following we consider a slightly modified version of the Hamiltonian
Eq.(1) of Ref.\cite{Greco1} describing the t-J model coupled to 
dispersionless phonons. First, we assume that the 
underlying internal symmetry is the symplectic group Sp(N/2) and not SU(N). 
Consequently, we split the internal index $p$ in Eq.(1) of Ref.\cite{Greco1}
into an index $\sigma$ describing the two spin components and a flavor
index $\mu$ running from 1 to N/2. Secondly, we modify the
second term in Eq.(1) in the following way:
\beq
\sum_{{ij}\atop{p,p'=1...N}} {J_{ij}\over{4N}}X_i^{pp'} X_j^{p'p}
\rightarrow
\sum_{{ij,\sigma \sigma'=1,2}\atop{\mu,\mu'=1...N/2}}
{J_{ij}\over{4N}}
X_i^{\mu \sigma,\mu \sigma'} X_j^{\mu' \sigma',\mu' \sigma}.
\label{J}
\eeq
In the modified term the sum over the flavor 
index is carried out independently in each X operator, whereas the sum over 
spins retains the old form. The two terms in 
Eq.(\ref{J}) are identical in the physical case N=2. At large N they
differ, however, and the label arrangement in the left and right terms 
in Eq.(\ref{J}) favor  RVB and antiferromagnetic correlations,
respectively. Thirdly, we multiply the Hamiltonian with an overall factor 2
so that it coincides with the usual Hamiltonian for $N=2$ and $t$ and $J$ 
have the usual meaning.

The 11 element of the electronic Green's function matrix in a  BCS
superconductor has the following form,
\beq
G(i\omega_n,{\bf k}) = {1 \over{i\omega_n -\epsilon({\bf k}) -
\Sigma(i\omega_n,{\bf k}) -\Delta^2({\bf k})/(i\omega_n + \epsilon({\bf k})
-\Sigma(i\omega_n,{\bf k}))}}.
\label{G11}
\eeq
$\omega_n=(2n+1)\pi T$ is a fermionic Matsubara frequency and $T$ 
the temperature. $\Delta({\bf k}) = \Delta(cos(k_x)-cos(k_y))$ 
describes a frequency-independent superconducting order
parameter with d-wave symmetry which is a rather good approximation
for the anomalous self-energy at large N\cite{Zeyher1}.
The one-particle energy $\epsilon({\bf k})$ is in the leading $O(1)$
of the 1/N expansion given by $-2t(cos(k_x)+cos(k_y)) -4t'cos(k_x)
cos(k_y)$. $t=t_0 \delta /2$ and $t'=t'_0 \delta /2$ are renormalized 
hopping elements with $t_0,t'_0$ being bare hopping matrix elements
between nearest and second-nearest neighbors on the square lattice,
respectively. $\delta$ is the doping away from half-filling. 
In our modified 1/N expansion 
$\epsilon({\bf k})$ does not have any contribution from $J$ as long as there
is no long-range antiferromagnetic order. $\Sigma$
is the diagonal self-energy due to the many-body interactions in $H$.
In the leading O(1/N) $\Sigma$ is additive in a purely electronic part
$\Sigma_{el}$ and a phononic part, $\Sigma_{ph}$, 
renormalized by electronic correlations.
We have analyzed all the contributions to $\Sigma_{el}$
and found that the RPA-like term
\beq
\Sigma_{RPA}(k) = -{T\over{NN_c}} \sum_{k_1}J({\bf k_1})
\gamma_s(k_1) G^{(0)}(k+k_1),
\label{RPA}
\eeq
with the spin vertex $\gamma_s(k_1) = 1/(1-J({\bf k_1}) a(k_1)/2)$,
is the only term which may give rise to interesting structures at 
low energies. In Eq.(\ref{RPA}) we combined the Matsubara
frequency
$i\omega_{n_1}$ and the two-dimensional wave vector $\bf k_1$ into a 
three-dimensional vector $k_1$. $N_c$ is the number of cells and $a(k_1)$ is  
defined below. The imaginary part of the remaining terms of 
O(1/N) exhibits  a featureless $\omega^2$ law at low energies  
and acquires structures only on the energy scales $J$ or $t$. This contribution
is similar to the total O(1/N) contribution to the imaginary part of $\Sigma$ 
in the usual 1/N expansion depicted in Fig. 5 of Ref.\cite{Greco}. 

The phonon contribution $\Sigma_{ph}$ is given in O(1/N) by
\beq
\Sigma_{Ph}(k) = {2T \over{N N_c}} \sum_{k_1} G^{(0)}(k+k_1)
{{16 \lambda \omega_o^2}\over{ (\omega_n-\omega_{n_1})^2 +\omega_0^2}}
\gamma_c(k_1+k,-k_1) \gamma_c(k,k_1),
\label{sigmaph}
\eeq
where $\omega_0$ and $\lambda$ denote the phonon frequency and the 
dimensionless electron-phonon coupling constant, respectively, and $G^{(0)}$
is given by Eq.(\ref{G11}) with $\Sigma = 0$. 
The phonon vertex $\gamma_c$ reads 
\beq
\gamma_c(k,k_1) = {{-1-b(k_1) + 2a(k_1)t({\bf k})} \over
{1+b(k_1)(1+c(k_1)) -a(k_1)d(k_1) +a(k_1)J({\bf k_1})/2}}.
\label{gammac}
\eeq
$t({\bf k})$ is equal to $-\epsilon({\bf k})/(\delta/2)$. 
The four susceptibilites $a,b,c,d$ are the generalizations of the 
normal state susceptibilities given in Eqs.(10)-(13) in Ref.\cite{Gehlhoff}
to the superconducting state.
Omitting the term proportional to $J$ in the denominator the expression
for $\gamma_c$ in Eq.(\ref{gammac}) is identical with the $J \rightarrow 0$ 
limit of the expression obtained in the 1/N expansion used in 
Ref.\cite{Zeyher}.


Fig. 1 shows the imaginary part of the electronic self-energy in the
superconducting state at $T=0$ as a function of the frequency $\omega$
for two momenta $\bf Q$ on the Fermi surface, namely, 
${\bf Q} = (1.33,1.33) = L$
along the diagonal (left panel) and ${\bf Q} = (3.14,0.26) = X$ 
(right panel). The energy unit is an effective nearest neighbor 
hopping constant $t$ of 150 meV. $t'/t$ is equal to -0.25 and the value
for $J$, $J=1.11$, is chosen such it reproduces the resonance peak in the 
spin susceptibility at
the experimental value of about 40 meV. The doping $\delta$ away from
half-filling is $\delta = 0.20$.  
The definition of the coupling constant $\lambda$, Eq.(7) of 
Ref.\cite{Greco1}, is based on the bare coupling and an average density of 
states of $1/8t$. The dashed and solid lines have been calculated 
neglecting and including vertex corrections, respectively. The dash-dotted 
lines represent the spin fluctuation contribution, Eq.(\ref{RPA}).

\begin{figure}[h]
      \centerline{\hbox{
      \epsfysize=8cm
      \epsfxsize=8cm
      \epsffile{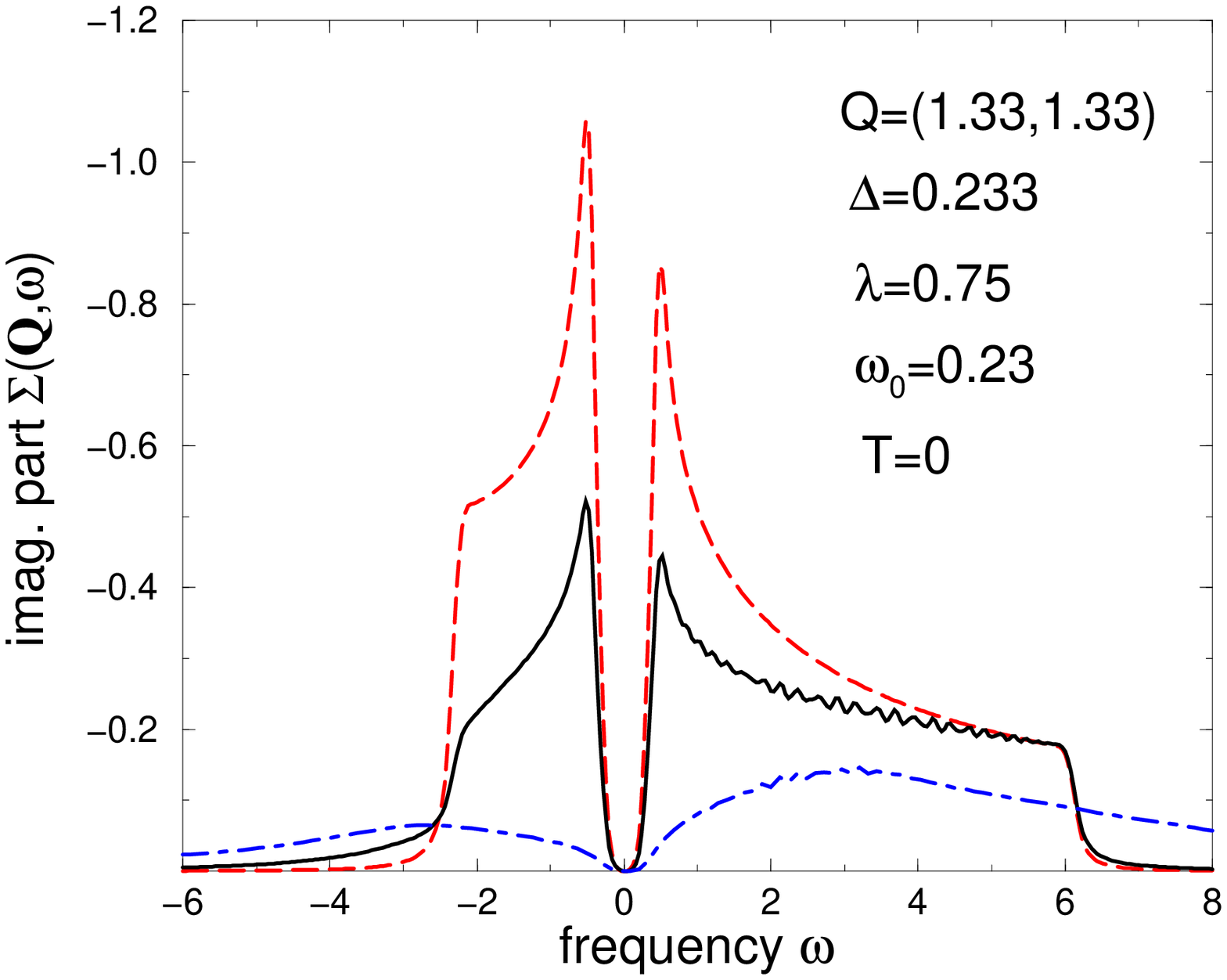}
      \epsfysize=8cm
      \epsfxsize=8cm
      \epsffile{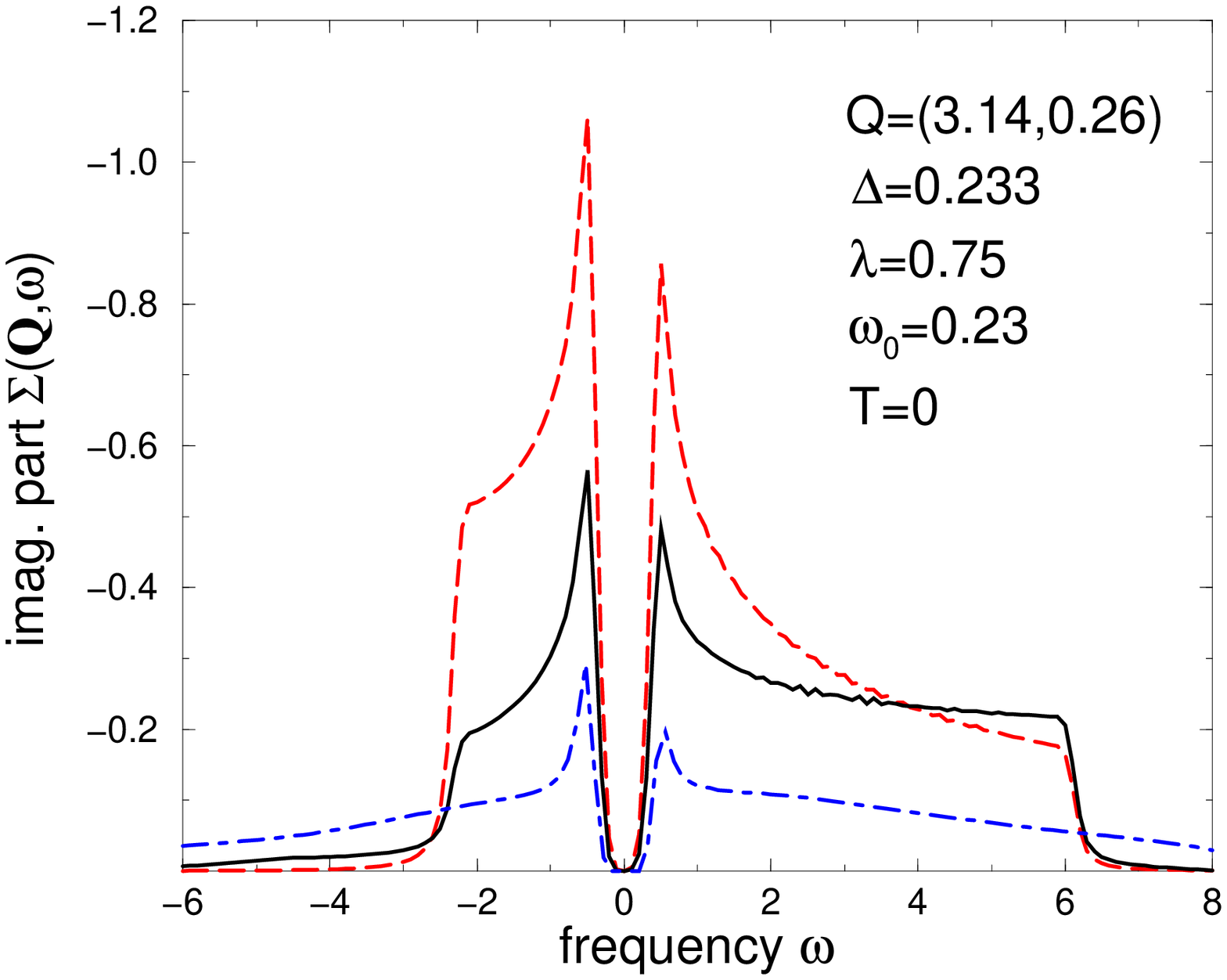}}}
\caption 
{Imaginary part of the electronic self-energy $\Sigma({\bf Q},\omega)$
at the points L (left panel) and X (right panel). Solid, dashed,
and dot-dashed lines correspond to the phonon contributions with and without
vertex corrections, and the spin fluctuation contribution, respectively.
Energies are given in units of $t=150$ meV.}
\label{fig1}
\end{figure}

The solid and dashed lines in Fig. 1 show a gap near the Fermi energy 
bounded by two sharp peaks with an energy distance of about 
$2(\Delta + \omega_0)$. The lines fall to zero below $\omega \sim -2.5$ and
above $\omega \sim 6$ reflecting the finite electronic band-width.
Comparing the left and right panel one recognizes that $\Sigma_{ph}$
is rather isotropic in spite of the fact that the vertex corrections 
depend strongly on momentum and that the superconducting gap is anisotropic.
The vertex corrections $\gamma_c$, which represent one effect of electronic
correlations, suppress large momentum transfers in the sum
over $\bf k_1$ in Eq.(\ref{sigmaph}) but enhance small momentum transfers.
Comparison of the dashed and solid lines shows that altogether the suppression 
prevails making the solid curves smaller by roughly a factor 2 compared to 
the dashed ones.

In order to discuss $\Sigma_{RPA}$ let us first consider the momentum
integrated susceptibility $\chi''(\omega)$ at large N, 
shown in the left panel of Fig. 2. In accordance with previous work 
$\chi''({\bf Q},\omega)$ develops a bound state in a very localized region 
around ${\bf Q} = (\pi,\pi)$ 
with nearly no spectral weight left at higher energies at this momentum.
Fig. 2 shows that also $\chi''(\omega)$ exhibits a sharp structure 
in the
region around 40 meV. It consists of a stronger component
at lower energy which comes from transitions near $(\pi,\pi)$ and 
corresponds to the 
resonance peak, i.e., an approximate  pole in the denominator of 
$\gamma_s$. The upper weaker peak is mainly due to
momenta near the X point and is caused by spin flip excitations across the
superconducting gap. Experimentally, a two peak structure has been
seen in inelastic neutron scattering in the momentum integrated spin
susceptibility in $YBa_2Cu_3O_{6.5}$ and $YBa_2Cu_3O_{6.7}$ (Figs. 5 and 6
of Ref.\cite{Fong}) but not in 
$\chi''({\bf Q},\omega)$ for ${\bf Q} = (\pi,\pi)$ in $YBa_2Cu_3O_{6.7}$
(Fig. 7 of Ref.\cite{Fong}). The
higher peak is sensitive to temperature, 
in contrast to the lower peak, and vanishes somewhere above the
superconducting transition temperature $T_c$. Guided by the result of
our calculation we identify the upper
peak in the experimental $\chi"(\omega)$ with the pseudogap with energies of 
about 60 and 55 meV in 
$YBa_2Cu_3O_{6.5}$ and $YBa_2Cu_3O_{6.7}$, respectively. Accepting this
interpretation the pseudogap increases slightly with decreasing doping
which is in line with the well-investigated behavior of the pseudogap
in the $Bi$-cuprates. Fig. 2 also illustrates that $\chi''(\omega)$
extends over a large energy region set by the band-width and that most of 
the spectral weight resides at high energies. Using the exact sum rule
one finds that the resonance peak exhausts only about 1.5 per cents of the 
total spectral weight which agrees well with experiment\cite{Aeppli,Fong2}. 
\begin{figure}
      \centerline{\hbox{
      \epsfysize=8cm
      \epsfxsize=8cm
      \epsffile{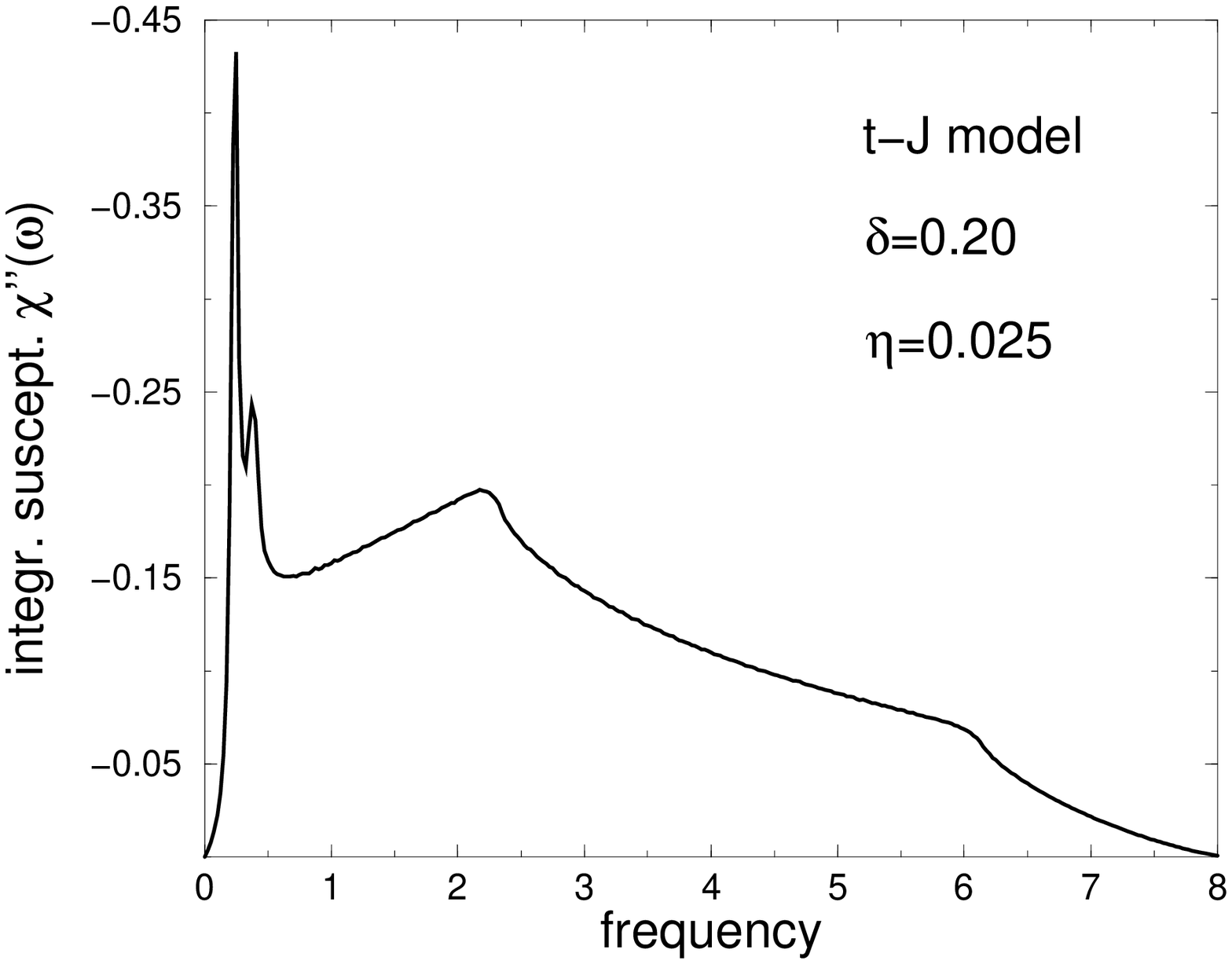}
      \epsfysize=8cm
      \epsfxsize=8cm
      \epsffile{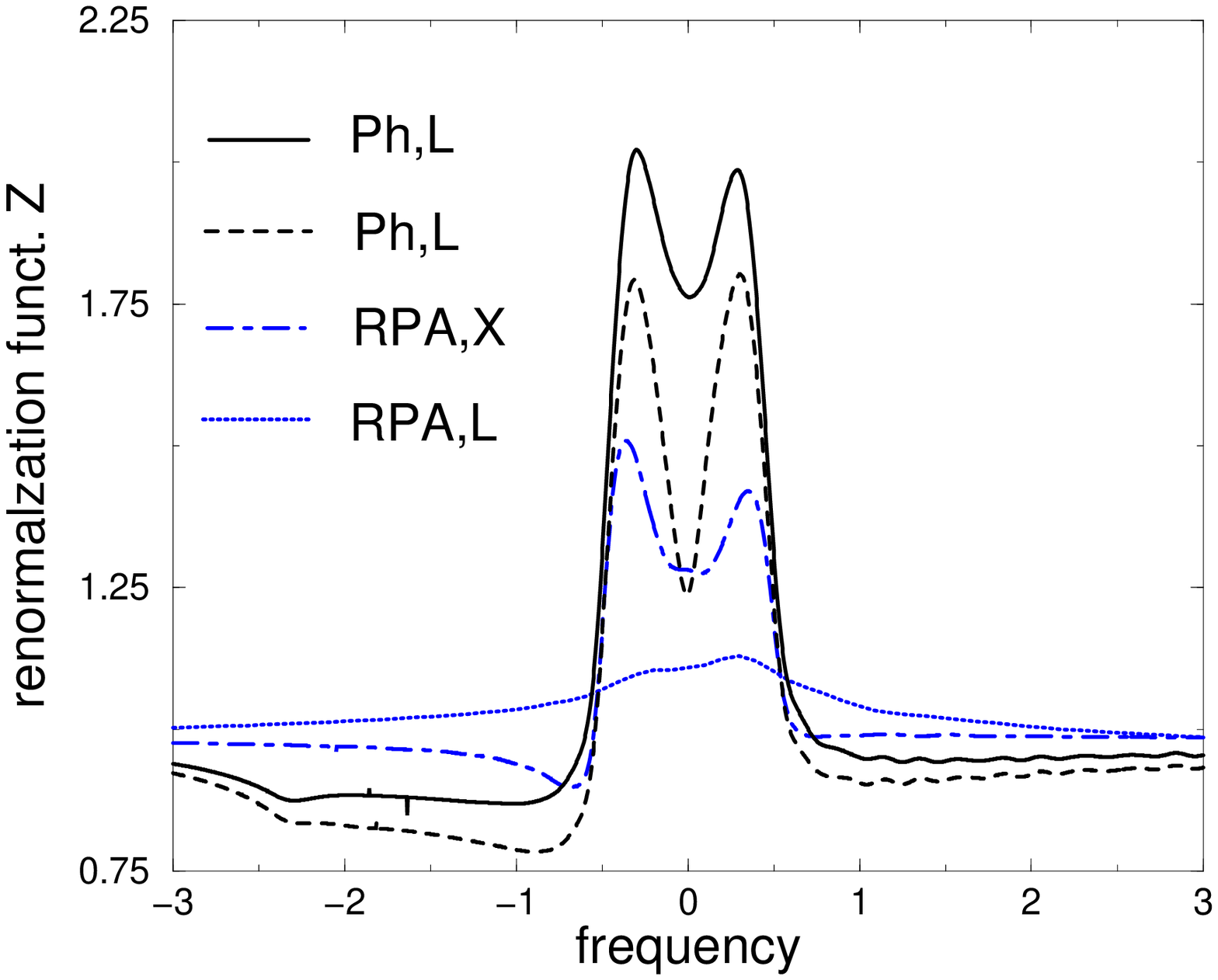}}}
\caption 
{Left panel: Momentum-integrated susceptibility $\chi''(\omega + i\eta)$ 
using the same parameters as in Fig. 1; right panel: renormalization function
$Z({\bf Q},\omega)$ for interaction with phonons (``Ph'') at the point 
${\bf Q}= L$ for $T=0$ (solid line), $T=180K$ (dashed line),
and with spin excitations (``RPA'') at ${\bf Q}=L$ and $T=0$
(dot-dashed line) and at ${\bf Q}=X$ and $T=0$ (dotted line).}
\label{fig2}
\end{figure}

The dash-dotted lines in Fig. 1 represent the imaginary part of $\Sigma_{RPA}$.
In contrast to $\Sigma_{ph}$ $\Sigma_{RPA}$ is strongly anisotropic along
the Fermi surface. The low-frequency spin fluctuations which are concentrated
around momentum transfers of about $(\pi,\pi)$, can probe the superconducting
gap if $\bf Q$ is near the X but not the L point. Therefore sharp gap 
features appear only in the right but not in the left panel. One also
notes that the dot-dashed lines are substantially smaller in absolute value
at low energies and decay slower towards higher energies compared to the 
case of phonon scattering. The reason for this behavior can be inferred
from $\chi''(\omega)$, as shown in the left panel of Fig. 2. Most of the
spectral weight in $\chi''(\omega)$ is found at high energies which leads
to the slow decay of the dash-dotted curves towards high energies. The
rather small spectral weight associated with low energies and the resonance
peak yields, together with $J$ as the coupling constant between electrons and
spin fluctuations, only small values in the self-energy near the Fermi 
energy. The neutron resonance peak leads in the calculation of 
Ref.\cite{Eschrig} to much larger effects in $\Sigma$ and $Z$. There the 
employed coupling constant and the frequency-integrated resonance peak
are taken as 0.65 eV and 13 per cents of the total sum rule, 
respectively. The corresponding values are in our case $J=0.166eV$ and 
1.5 per cents, respectively. Our rather small spectral weight of the resonance
peak agrees well with the experiment\cite{Aeppli,Fong2}. The present 1/N
expansion also identifies in a unique way $J$ as the coupling constant
between electrons and spin excitations.

The right panel in Fig. 2 shows the renormalization function 
$Z({\bf Q},\omega)=1-Re({\Sigma({\bf Q},\omega)-\Sigma({\bf Q},0)})/\omega$,
where $Re$ denotes the real part. The solid line is calculated from the 
phonon induced self-energy at the point L and at $T=0$. A very similar curve
is obtained for the point X. The solid line deviates substantially from 1
only in a narrow energy range around the Fermi energy set by the phonon
energy and the superconducting gap assuming a value of 1.76 at 
$\omega =0$. Defining a renormalized coupling function 
$\tilde{\lambda}({\bf Q})$
by $Z({\bf Q},0) = 1+\tilde{\lambda}({\bf Q})$ one finds a rather isotropic
$\tilde{\lambda}({\bf Q})$ of 0.76. There are two competing effects due
to electronic correlations in renormalizing the original bare coupling
constant $\lambda$: Vertex corrections suppress $\Sigma_{Ph}$ by roughly 
a factor 2, but density of state effects increase 
the effective $\lambda$ again so that $\tilde{\lambda}$ and $\lambda$ do not
differ much. Calculating the related quantitiy $\lambda_{tr}$,
which determines the coupling strength in transport phenomena, one has for
the bare values $\lambda_{tr} = \lambda$ because the electron-phonon
interaction was assumed to be momentum-independent. The renormalized quantity
$\tilde{\lambda}_{tr}$ is, however, equal to 0.24, i.e., it is about three 
times smaller than $\tilde{\lambda}$. In this case the strong suppression of 
large-momentum scattering by vertex corrections dominates. Such a big
difference between $\tilde{\lambda}$ and $\tilde{\lambda}_{tr}$ is needed to 
have agreement with experiment: The observed slope change in the quasi-particle
dispersion by about a factor 2 corresponds to $\tilde{\lambda} \sim 1$,
whereas the transport data suggest a three or four times smaller 
$\tilde{\lambda}_{tr}$ in the optimally doped case.

The dashed curve in the right panel of Fig. 2 describes the phonon-induced
$Z$ at the L point at T=180K. It illustrates that its value at $\omega = 0$
is sensitive to temperature and thus would decrease the slope change
substantially with increasing temperature. The dotted and dash-dotted
curves in Fig. 2 represent the spin fluctuation induced renormalization
factor. It is very anisotropic and in general much smaller than the
phonon-induced contribution to $Z$. As shown in Ref.\cite{Greco} the
transport quantity $\tilde{\lambda}_{tr}$ tends generally to be larger than
$\tilde{\lambda}$ in the case of spin fluctuations. The dotted and 
dash-dotted curves
thus yield an effective $\tilde{\lambda}_{tr}$ of about 0.2 which is compatible
with transport but indicates, on the other hand, that the spin fluctuation
contribution to $Z$ is not the dominant one. 

The left panel in Fig. 3 shows the evolution of the spectral function at 
low temperatures
in the superconducting state as a function of frequency for six momenta
approaching the Fermi wave vector in the direction L. The interaction with
phonons as well as the parameters of Fig. 1 were used in calculating the 
self-energy. The dashed lines represent the spectral function without
the self-energy. Going away from the Fermi surface in momentum space
the quasi-particle peak disperses much weaker than the free peak away
from the Fermi energy and looses rapidly spectral weight. The weak hump
seen near the Fermi surface disperses roughly with the free particle
dispersion at somewhat lower energy and becomes broad further away from
the Fermi surface. Plotting the same as a function of momentum for a
fixed energy one essentially obtains Lorentzians dispersing like free
particles. All these properties are well reflected in the experimental
data\cite{Bogdanov,Valla,Kaminski,Lanzara} showing that the self-energy
is rather isotropic along the Fermi surface but strongly frequency-dependent.
\begin{figure}
      \centerline{\hbox{
      \epsfysize=8cm
      \epsfxsize=10cm
      \epsffile{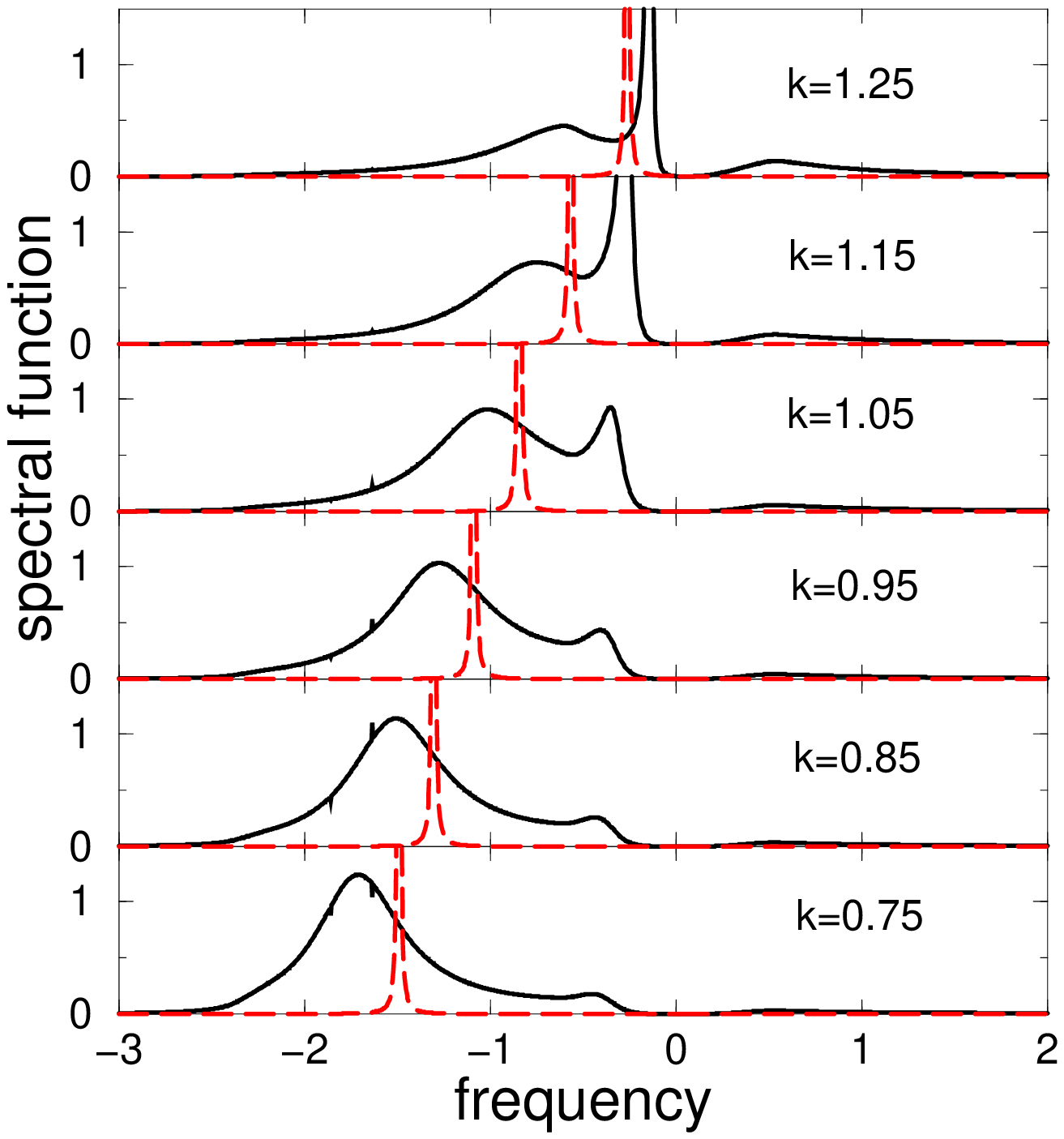}
      \epsfysize=8cm
      \epsfxsize=8cm
      \epsffile{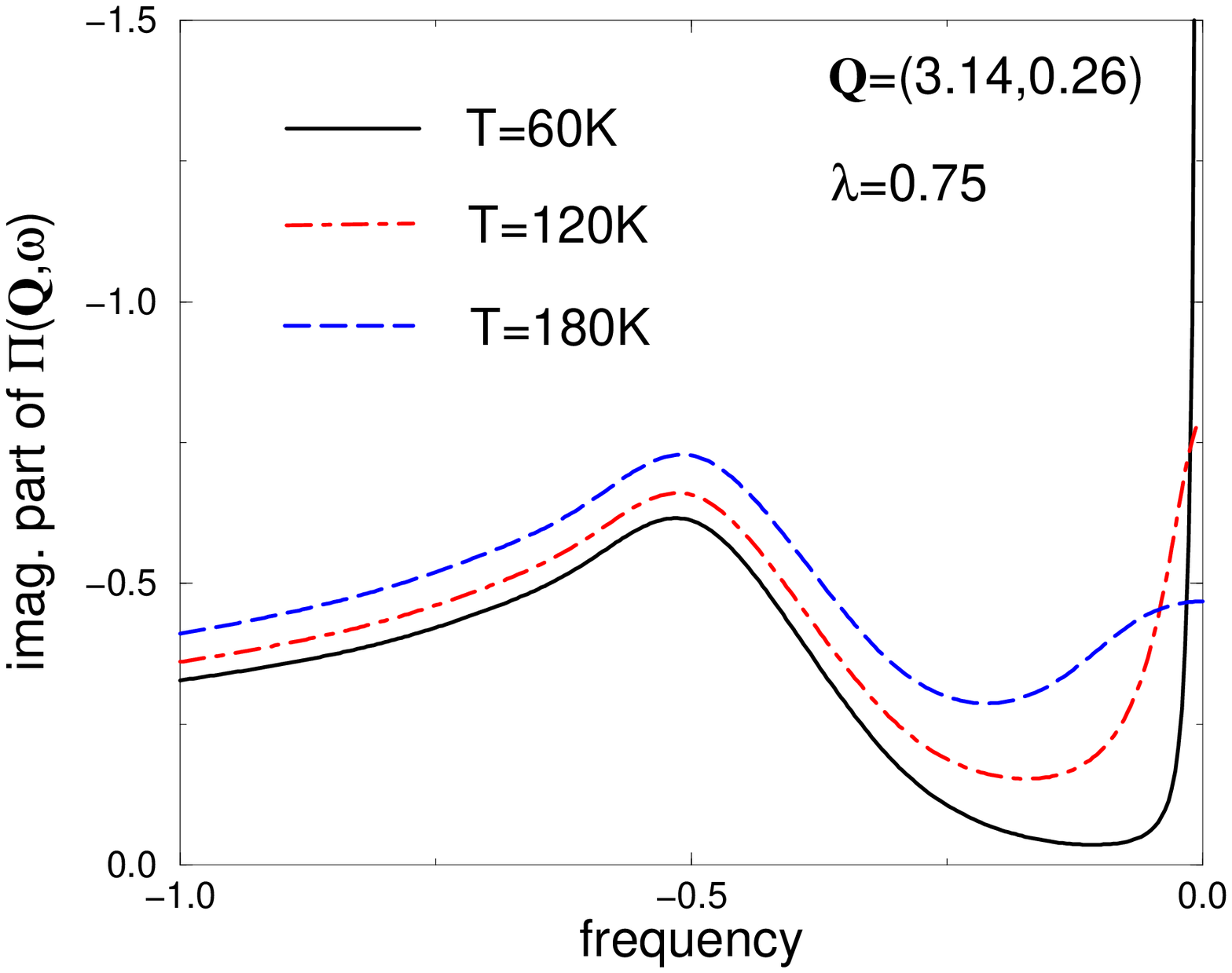}}}
\caption
{Left panel: Spectral function for phonon scattering for 6 momenta 
approaching the Fermi vector
along the $(\pi,\pi)-$direction, dashed curves are obtained neglecting 
self-energy effects; 
right panel: imaginary part of $\Pi({\bf Q},\omega)$ for three temperatures,
using parameter values as in Fig. 1.}
\label{fig3}
\end{figure}
From this one concludes again that the spin fluctuation contribution to $Z$
cannot dominate in agreement with the right panel in Fig. 2.

The right panel in Fig. 3 shows the imaginary part of the total self-energy
$\Pi$ associated with $G$, i.e., $\Pi(i\omega_n,{\bf k}) = 
- G^{-1}(i\omega_n,{\bf k})+i\omega_n -\epsilon({\bf k})$. The curves
in this Figure can directly be compared with the corresponding curves
extracted from ARPES data, Fig. 2a of Ref.\cite{Norman}. 
In the calculation we used for all temperatures the same T=0 gap $\Delta$
which approximately is found in the analysis of the data\cite{Norman}.
There is good agreement
on a qualitative level. Quantitatively, there are, however, several
discrepancies: in order to reproduce the absolute value of the hump
(our energy unit t corresponds to 150 meV) $\lambda$ should be increased
by about a factor 2 which would tend to yield too large slope changes
in the quasiparticle dispersion at the point L. At low temperatures
the gap between the hump and zero energy is in our case much more pronounced
than in the experiment. One reason for this discrepancy is our
use of just one phonon frequency instead of the true, rather broad 
phonon density of states. It is also possible that the spin fluctuation
contribution cannot be totally neglected at the point $X$ and enhance the
dump-dip feature. 

The situation is less clear in the underdoped region. Experimentally,
the momentum-integrated susceptibilities, if integrated between 0 and
65 meV, are typically one order of magnitude or more larger than in the 
optimally doped case. For instance, $YBa_2Cu_3O_{6.7}$ exhausts in this
region about 17 (10) per cents of the sum rule at T=12K (200K)\cite{Fong}
whereas
the resonance peak does only 1 per cent\cite{Aeppli,Fong2}. 
Using the experimental
susceptibility in Eq.(\ref{RPA}) instead of the calculated large N 
susceptibility we find in the above case 
for ${\bf Q} = L$ $Z({\bf Q},0) = 1.47 (1.27)$ and for 
${\bf Q}= X $ $Z({\bf Q},0) = 3.30 (2.31)$ at T=12K (200K).
Because $\lambda_{tr} \geq \lambda$ in the case of spin fluctuations
the resulting values for $\lambda_{tr}$ seem to be too large
to be compatible with the resistivity curves\cite{Ito}. Also the 
large anisotropy in the slope change of the quasi-particle dispersion,
implied by these values, seems not to be supported by the available 
experiments.  


Using a 1/N expansion which takes into account the magnetic resonance peak
already in leading order we have calculated the contributions of
O(1/N) to the electronic self-energy for a t-J-Holstein model. 
We find that the recently observed 
low-energy renormalization of quasi-particles in the superconducting
state of cuprates can well accounted for taking a phonon energy of 35 meV 
and a dimensionless coupling constant of about $0.75 - 1$.
In calculating the spin fluctuation part to the self-energy our approach
does not contain any adjustable parameter. Due to the rather small 
value of $J$, which acts as the coupling constant, and due to the 
structureless $\chi''$ at large N we find that this contribution is rather
small and strongly anisotropic along the Fermi surface. This means that it
does not play the dominant role in the observed low-frequency renormalization
of quasi-particles, at least, in the optimal and overdoped regime.
\vspace{-0.5cm}

\end{normalsize}
\end{document}